# Predicted photonic band gaps in diamond-lattice crystals built from silicon truncated tetrahedrons

Léon A. Woldering,[1, *] Leon Abelmann,[1] and Miko C. Elwenspoek[1, 2]

[1]*Transducers Science and Technology (TST),*
*MESA+ Institute for Nanotechnology,*
*University of Twente, P.O. Box 217,*
*7500 AE Enschede, The Netherlands*

[2]*FRIAS, University of Freiburg, Albertstrasse 19, D-79194 Freiburg, Germany*

## Abstract

Recently, a silicon micromachining method to produce tetrahedral silicon particles was discovered. In this report we determine, using band structure calculations, the optical properties of diamond-lattice photonic crystals when assembled from such particles. We show that crystal structures built from silicon tetrahedra are expected to display small stop gaps. Wide photonic band gaps appear when truncated tetrahedral particles are used to build the photonic crystals. With truncated tetrahedral particles a band gap with a width of 23.6% can be achieved, which is more than twice as wide compared to band gaps in self-assembled diamond-lattices of hard-spheres. The width of the band gap is insensitive to small deviations from the optimal amount of truncation. This work paves the way to a novel class of silicon diamond-lattice band gap crystals that can be obtained through self-assembly. Such a self-assembly approach would allow for easy integration of these highly photonic crystals in existing silicon microfluidic and -electronic systems.

## I. INTRODUCTION

There is substantial interest in the self-assembly of particles with different shapes and sizes to obtain functional structures.[1–9] In particular there is a lot of attention for the self-assembly of particles into periodic structures with a diamond crystal lattice because of the expected interesting properties of these structures.[10–12] For instance, many efforts are devoted to obtaining diamond-lattice crystals that have a photonic band gap. When a crystal has a photonic band gap,[13,14] it means that there is a range of frequencies of light that can not exist in the structure. This phenomenon allows for many interesting possible applications, such as thresholdless lasers,[15,16] applications in sensing,[17,18] and control of the spontaneous emission of embedded light sources.[19–21] In a few studies, see for example references[22–24], it was suggested that particles with tetrahedral symmetry can self-assemble in periodic crystal structures with a cubic diamond lattice.

A few ways have been demonstrated to obtain tetrahedral particles made from silicon.[25,26] In a previous article by our group silicon tetrahedral particles were described that are obtained by a fabrication scheme that utilizes silicon micromachining techniques, which makes them readily available.[27] The size of the obtained particles can be varied and more importantly, the employed fabrication techniques allow for additional functionalization of the tetrahedron vertices, faces and edges. These functionalization options are beyond what is demonstrated or feasible with the other types of silicon tetrahedral particles and with other particles in general. This potential for additional and extended functionalization makes these structures interesting platforms to study. For instance, it was suggested to micromachine tetrahedral particles which have conducting split-rings for the benefit of negative index of refraction materials.[28]

In this report we demonstrate the expected optical properties of diamond lattice crystals assembled from silicon tetrahedra. The photonic properties are quantified using band structure calculations. Furthermore, we show that interesting results are obtained for diamond-lattice crystals from a truncated type of silicon tetrahedral particles. We calculate that such periodic structures have photonic band gaps with widths larger than 23%. Our results provide insight in the amount of truncation necessary to obtain an optimal photonic band gap when assembling these crystal structures.

The self-assembly of silicon (truncated) tetrahedral particles is an interesting field of study in itself. By varying the fraction of truncation of the particles, extended studies into particular aspects of self-assembly are possible. In particular, by varying the truncation fraction it is possible to study in detail how the self-assembly depends on the inter-particle contact area. More truncation allows for larger contact areas between the individual particles, thus increasing the expected inter-particle binding forces. In addition, the stability of the crystals benefits from the resulting larger contact area. Since the results that will be described here also indicate the expected reflectivity from these crystals, reflectivity measurements can in the future be compared to the calculated results to determine the quality of obtained crystals. This is briefly explained in section IV A.

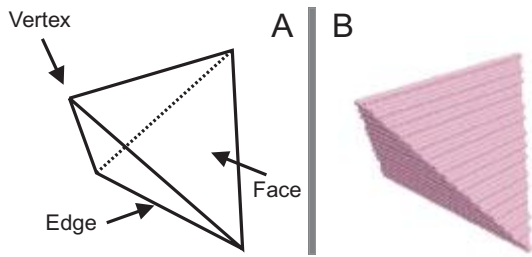

FIG. 1: (Color online) A) Schematic illustration of a tetrahedron. These pyramid-like structures have four faces, four vertices, and six edges. B) Ray-traced image of a space-filled tetrahedron. The tetrahedron is assembled from spheres that have their centers inside the structure and on the faces. The radii of the space-filling spheres are chosen to be slightly larger than strictly necessary to avoid interfacing issues. In this case the structure is defined by 165924 spherical elements.

The actual self-assembly of these particles is a work in progress and is beyond the scope of this paper. Nonetheless, we want to point out that the mentioned truncation of the silicon tetrahedral particles can be achieved in a fashion that is complementary to the fabrication of the particles themselves.[27] The extra contact surfaces that arise because of truncation can be chemically functionalized to assist the self-assembly process.

## II. DEFINING THE CRYSTAL GEOMETRY

A tetrahedron is a polyhedron that is also known as a triangular pyramid. This pyramid has four faces, four vertices and six edges, see Figure 1 for a schematic illustration (A) and a ray-traced image (B) of a tetrahedron. When assembled, the tetrahedra are in contact at the tips to form a crystal structure with a diamond lattice. Figure 2 is an illustration of a section of a crystal consisting of 13 assembled tetrahedral particles. As described below, all tetrahedra used in the described calculations are generated using a space-filling method using small spheres. For clarity and completeness we will show actually used space-filled geometries in most of the images in this report. The images are generated by a ray-tracing program and have a slight perspective view.

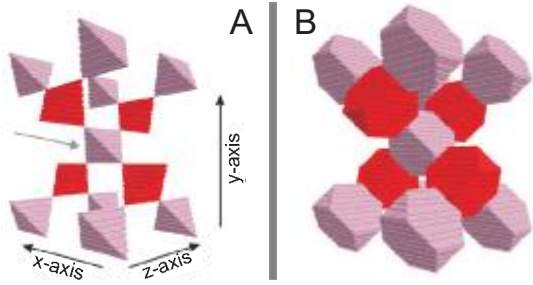

FIG. 2: (Color online) A) Ray-traced image of a part of a crystal consisting of assembled tetrahedra in a diamond lattice. The section that is shown consists of 13 particles. In a tetrahedral diamond crystal structure, all tetrahedrons are in contact with other tetrahedra at the four vertices. This condition can be observed for the central tetrahedron indicated by the grey arrow. Furthermore, the $x$-, $y$-, and $z$-axis are defined. The two colors indicate identical tetrahedra with two different orientations. B) Similar image of a section consisting of 13 particles that are truncated. The particles are oriented in an orthorhombic lattice with the same lattice spacings as in A). It can be seen that the filling fraction of tetrahedron material in this structure is significantly higher when compared to crystals from tetrahedra which are not truncated.

Band structure calculations are performed with the MIT photonic bands package, referred to as "$mpb$". This package solves Maxwell's equations in infinitely periodic structures using a numerical, iterative procedure.[29] The required geometry is entered into the program after which $mpb$ maps the distribution of the dielectric constants on a three-dimensional grid. The calculation algorithm is fully vectorial. Typically generated output are the eigenstates and eigenvalues of the periodic structure. It is also possible to obtain, for instance, electric field distributions throughout the defined three-dimensional grid. There is no function that defines tetrahedra in this package. For this reason we chose to use a space-filling method with spheres to define the tetrahedra. Furthermore, by using this space-filling method it is possible to define truncated tetrahedra in a straightforward fashion.

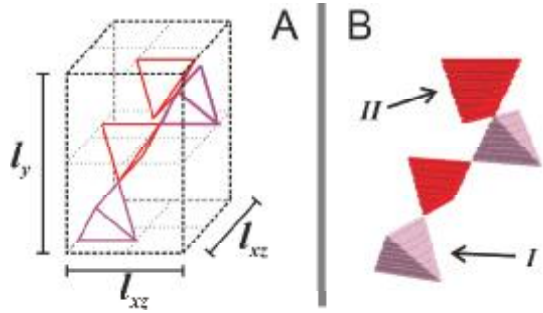

FIG. 3: (Color online) A) Schematic illustration of the minimum geometry that was used to define the periodic structures in *mpb*. The crystal has a orthorhombic lattice, which is defined by long lattice spacing $l_y$ and two short lattice spacings $l_x$ and $l_z$. The two short lattice spacings are equal because of symmetry reasons and are therefore both indicated as $l_{xz}$. Four tetrahedra are required to build this geometry, two of type $I$ and two of type $II$. Different colors are used to distinguish them. These two types of tetrahedra are identical and differ only in orientation. B) Ray-traced image of the unit-cell. Tetrahedra types $I$ and of $II$ are indicated. Multiplying this unit-cell along the three lattice spacings yields the infinitely periodic structure that *mpb* uses for its calculations.

Although the crystals we study have a cubic unit-cell, it is more straightforward to analyze them using an orthorhombic unit-cell, similar to what was done for inverse woodpile photonic crystals.[30,31] The orthorhombic layout of the minimum required geometry that was used to define the crystal structure is depicted in Figure 3. From here on we will refer to this minimum geometry as the unit-cell. The unit-cell has three lattice spacings. The long lattice spacing $l_y$ is defined parallel to the $y$-axis, and the short lattice spacings $l_x$ and $l_z$ are parallel to the $x$- and $z$-axes respectively. The lengths of these lattice spacings are related as follows:

$$l_x = \frac{l_y}{\sqrt{2}} \text{ and } l_x = l_z. \tag{1}$$

Since $l_x$ and $l_z$ can be considered equal due to symmetry reasons, the terminology is reduced to two lattice spacings, $l_y$ and $l_{xz}$, which are indicated in Figure 3(A). The Figure also shows that the periodic structure is built from identical tetrahedral particles with two different orientations. These tetrahedra are referred to as type $I$ and $II$. They can also be identified in Figure 2.

Figure 3(B) shows a ray-traced image of the unit-cell; the minimum required geometry that was used to define the crystal structure. With this minimum geometry, *mpb* builds an infinitely periodic structure in three dimensions by multiplying the unit-cell along the three lattice spacings. A script was written to generate the unit-cell according to the procedure outlined in appendix A. To verify that the obtained space-filled geometries are correct, typical results were rendered with the ray-tracing program and inspected from various angles, see for instance Figures 1, 2, and 6.

### III. TWO IMPORTANT RESOLUTIONS

A standard desktop computer with a 3GHz dual-core processor and 2Gb of internal memory was used for the *mpb*-calculations. For these calculations a number of resolutions and settings are of importance. Two of these will be described here. The first is the resolution with which the three-dimensional grid is defined, the so-called grid resolution. This resolution is set at 68 x 96 x 68, along the $x$-, $y$-, and $z$-directions respectively. Further increasing the grid-resolution is restricted by the amount of available memory. Nevertheless, we will show that this setting is sufficient for these calculations.

The second important resolution is the density of the space-filling spheres that define the geometry, which we will refer to as "sphere-density". Since spheres are generated with their centers inside and on the faces of the tetrahedra, the volume of the tetrahedra is overestimated at low sphere-densities. When the set density of spheres goes to infinity, the radius goes to zero and the tetrahedra should be nearly perfectly defined. To check the geometry and verify that the grid- and sphere-densities are sufficiently high, we performed *mpb*-calculations for a range of sphere-densities and plotted the obtained *mpb*-calculated filling fraction of silicon as a function of the set sphere-density, see Figure 4. In this plot the sphere-density is represented by the amount of elements used along lattice spacing $l_y$.

The data show that as the density is increased, the *mpb*-calculated volume fraction of silicon decreases as is expected. Since the volume fraction is expected to decrease asymptotically to the analytically calculated value, the data was fitted to the following asymptotic equation:

$$\phi_{\text{Si}} = a + \frac{b}{\text{density}^c}, \qquad (2)$$

with $a$, $b$, and $c$ free parameters. Constant $a$ is the limit of the asymptotic expression and should be equal to the analytically calculated value for the volume fraction of silicon $\phi_{\text{Si}}$ in the geometry, see appendix B. Parameter $b$ depends on the chosen $x$-axis units and is irrelevant here. Parameter $c$ depends on the change in volume induced by the decreasing radii of the space-filling spheres. The value of this parameter is expected to be just over 1.

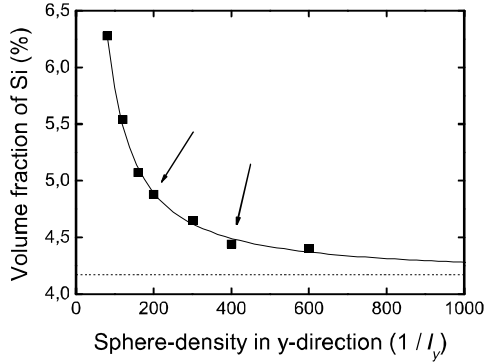

FIG. 4: $Mpb$-calculated volume fraction of silicon $\phi_{Si}$ as a function of set sphere-density, represented by the amount of spherical elements used along lattice spacing $l_y$. Upon increasing density, the calculated volume fraction asymptotically decreases towards $\phi_{Si} = 4.17 \pm 0.08\%$, which is in agreement with the analytically calculated value for the volume fraction of silicon in the unit-cell of $\phi_{Si} = 4.17\%$. The black curve is the best fit of the used asymptotic expression to the data. This result verifies that the geometry of the crystal structures is correctly defined and that the grid-resolution is high enough. The arrows indicate the sphere-densities that were mainly used in this work. Higher densities could not be used because of too long calculation times.

Fitting equation 2 to the data in Figure 4 yields the following result: $a = 4.17 \pm 0.08$, which is in excellent agreement with the analytically calculated volume fraction of $\phi_{Si} = 4.17$. Furthermore, $c = 1.2 \pm 0.11$, which is indeed close to one. The correlation coefficient is excellent ($R^2 = 0.997$). This result shows that the grid-resolution used in the calculations is high enough and that the crystal geometry is correctly defined.

No data was calculated for sphere-densities exceeding 600 $(1/l_y)$. The reason for this is that the initialization of calculations with densities of 600 $(1/l_y)$ already took many days. Increasing the density to larger values was deemed to be impractical, certainly in comparison with the relatively low achieved improvement in calculated volume fractions. Consequently, we have chosen to use sphere-densities of 400 $(1/l_y)$ for the calculations.

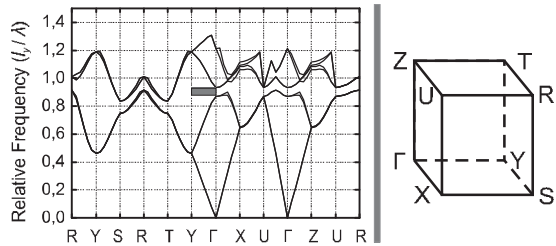

FIG. 5: Left: *Mpb*-calculated bandstructure of a crystal with a diamond lattice built from untruncated tetrahedra. The tetrahedra have a dielectric constant of $\varepsilon_{Si} = 12.25$. There is no apparent photonic band gap. In grey a small stop gap is indicated for the $\Gamma$-Y direction. This stop gap has a width of $(\frac{\Delta\omega}{\omega_{gap}}) = 3.7\%$ around frequency $\omega_{gap} = 0.901$ $(l_y/\lambda)$. Right: a schematic of the Brillouin zone of the orthorhombic lattice. In total 8 symmetry points are indicated. Of these, the $\Gamma$, X, Y, and Z points are of particular interest since the $\Gamma$-X, $\Gamma$-Y, and $\Gamma$-Z directions are parallel to the real-space $x$-, $y$-, and $z$-axes respectively. These real-space directions can be easily identified when observing fabricated crystals using microscopy techniques.

At this sphere-density the calculated volume fraction of silicon is overestimated by only 0.27%point, which is sufficiently precise for our purposes. When calculating the properties of periodic structures built from truncated tetrahedra, we chose a sphere-density of 200 $(1/l_y)$ to obtain initial results and performed more precise calculations with a sphere-density of 400 $(1/l_y)$ at interesting truncation fractions. This density is more than 64 times ($4^3$) higher compared to the grid-resolution used in *mpb*.

## IV. RESULTS AND DISCUSSION

### A. Calculated optical properties of a crystal built from perfect tetrahedra

Figure 5 shows the calculated band structure for a periodic crystal consisting of untruncated, normal tetrahedra. In our calculations bands 1 to 8 were evaluated with 72 $k$-points. The chosen tetrahedron material is silicon, for which a dielectric constant of $\varepsilon_{Si} = 12.25$ was set. This dielectric constant was chosen based on a target wavelength of $\lambda = 1550$ nm. This wavelength is important for optical communication, and therefore of immediate interest for potential applications. In Figure 5 several small stop gaps can be identified. For example, a stop gap is found in the $\Gamma$-Y direction, indicated by the grey bar. Since the $\Gamma$-Y direction is parallel to the real-space $y$-axis, a small peak is expected to appear when performing optical reflectivity experiments along that direction. Reflectivity measurements can in the future be compared to the calculated results to determine the quality of obtained crystals. This is explained as follows. The reflectivity of photonic crystals is due to Bragg-diffraction,[32] which occurs when there is long range periodicity. Therefore, multiple unit-cells are probed when performing reflectivity experiments on photonic crystals.[33] Consequently, results of optical reflectivity experiments are indicative for the interior quality of the self-assembly process. Thus, reflectivity results combined with scanning electron microscopy allow for the interior and the exterior of the geometry to be probed. By using these techniques the self-assembly process can relatively easily be qualified and optimized.

The calculated stop gap has a width of $(\frac{\Delta\omega}{\omega_{\text{gap}}}) = 3.7\%$ around a central frequency of $\omega_{\text{gap}} = 0.901$ $(l_y/\lambda)$. This $\omega_{\text{gap}}$ is a relative frequency and multiplication with a real-space $l_y$, as determined by the geometry of obtained crystals, yields the absolute frequencies for the calculated gaps. When calculated for the target telecommunication wavelength $\lambda = 1550$ nm, $l_y$ equals 1397 nm.

The resulting band structure in Figure 5 indicates that this crystal is not expected to display photonic band gap behavior, *i.e.*, there is no range of forbidden frequencies of light. It can be surmised that this is due to the relative low amount of high refractive index material. To compare, for diamond structured photonic crystals obtained through various processes, typical optimal filling fractions of around 20% and more are described.[10,34] Here, the filling fraction is only $\phi_{\text{Si}} = 4.17\%$, which is well below the described optimal values. This result also implies that a photonic band gap cannot be expected in diamond-lattice crystals assembled from tetrahedra made from materials with lower dielectric constants, for example polymers and silicon nitride.

By assembling tetrahedra that are truncated, photonic crystals with higher filling fractions can be obtained. In the following sections we will analyze what the expected photonic crystal properties are of such periodic structures. Furthermore, we will determine what the optimal truncation fraction $t$ is to obtain a photonic crystal with a maximized width of the band gap $(\frac{\Delta\omega}{\omega_{\text{gap}}})$.

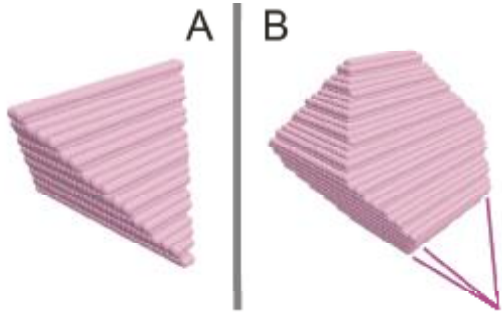

FIG. 6: Ray-traced images of A) an untruncated tetrahedron and B) a truncated tetrahedron. When the particle is truncated, smaller tetrahedron shaped tips are removed at the vertices. One of the removed tips has been sketched in. The truncating faces are parallel to the four original faces. Both particles were obtained with a sphere-density of 200 ($1/l_y$). They are not displayed at the same scale.

### B. Truncation

When a tetrahedron is truncated, in essence smaller tetrahedra are removed at its vertices. This removal is illustrated by the ray-traced image of the particles in Figure 6. On the left side (A) an untruncated tetrahedron is shown. When this pyramid is compared with the truncated tetrahedron in Figure 6(B), it shows that there are smaller tetrahedra missing at the tips. To support this explanation, the missing tip on the bottom right of Figure 6(B) has been sketched in.

Here the truncation is expressed as a fraction of lattice spacing $l_y$. When truncated tetrahedra are defined with the space-filling script, the size of the removed tetrahedral tips is increased by increasing the truncation fraction $t$, see Figure 7(A). This works as follows: by setting a truncation fraction $t$, the 4 planes which in the space-filling script define the top and bottom of the tetrahedra are shifted up and down along the $y$-axis. The planes are shifted with an amount $t$ such that the centers of the particles remain at the same position and the tetrahedra increase in size. However, there are also 4 truncating planes defined at the original vertices coordinates. Thus, as the tetrahedra are increased in size, tetrahedron shaped parts are removed at the vertices. With increasing truncation fractions $t$, the particles are truncated more significantly. In addition, the volume fraction of silicon $\phi_{\text{Si}}$ is increased because the particles get larger, see the calculated data in Figure 8. In the crystal the separate truncated tetrahedra are in contact at the truncating faces such that their overlap is maximized, see Figure 7(B) and Figure 2. The contact area $A_c$ between neighboring particles increases with increasing truncation fraction $t$.

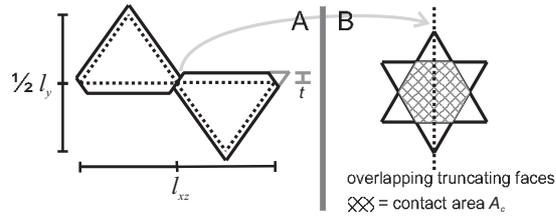

FIG. 7: A) Schematic of a cross-section of two untruncated (dotted lines) and truncated tetrahedra (solid lines) that are in contact. By setting a truncation fraction $t$ (indicated at the right vertex), the tetrahedra are increased in size, but remain at their original positions. At the original vertices, truncating planes were defined that remove smaller tetrahedra from the tips. In this way larger truncated particles are obtained that are in contact at the new faces. Note that at the top and bottom no truncation is apparent because there the cross section goes through the center of tetrahedron edges instead of through a vertex. Although the truncated tetrahedra appear to be misaligned, this is not the case, which can be observed in B). The dotted line indicates the contact as shown in A). With this alignment, the contact area $A_c$ (raster) between the two particles is maximized.

The truncated tetrahedra have 8 faces, 12 vertices, and 18 edges. There is a special condition when all the edges of the truncated tetrahedron have the same length. This is the polyhedron that in geometry is known as a truncated tetrahedron. Because we use the term truncated tetrahedron in a more general fashion, we will refer to this polyhedron as "perfect truncated tetrahedron". This particle is obtained at a truncation fraction of $t = \frac{1}{10}l_y$. A brief geometric analysis shows that for a "perfect truncated tetrahedron" the relation between the contact area $A_c$ and the lattice spacing $l_y$ is as follows:

$$A_c = \frac{3\sqrt{3}}{400} l_y{}^2. \qquad (3)$$

The structure in Figure 2(B) consists of 13 "perfect truncated tetrahedra".

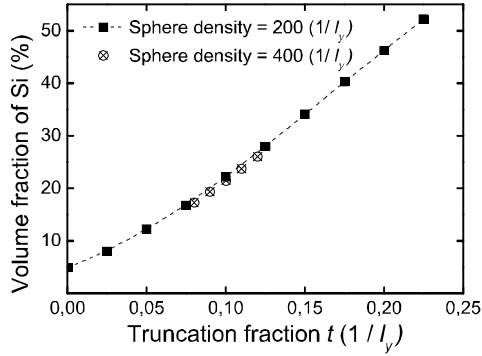

FIG. 8: Calculated volume fraction of silicon $\phi_{Si}$ versus truncation fraction $t$. These volume fractions are calculated with sphere-densities of 200 $(1/l_y)$ (black squares) and of 400 $(1/l_y)$ (crossed dots). With increasing truncation fraction, the filling fraction is increased because the particles get larger. At $t = 0.1$, the volume fraction obtained from our most precise calculation with a sphere-density of 400 $(1/l_y)$ is $\phi_{Si} = 21.47\%$. This value is only 0.77%point higher than the analytically calculated value for $\phi_{Si}$. The dashed line is a guide to the eye.

## C. Band gap properties versus truncation

To obtain a higher volume fraction of tetrahedron material in the crystal lattice, the truncation fraction $t$ is increased. Figure 8 shows the *mpb*-calculated volume fractions versus truncation fraction $t$. It can be seen that very high volume fractions can be obtained. As mentioned above, it is around $\phi_{Si} \approx 20\%$ that a maximum width of the band gap is expected. Here, volume fractions up to $\phi_{Si} \approx 50\%$ were calculated. Therefore it is apparent that the relevant range of volume fractions can be obtained when assembling truncated tetrahedra in a diamond lattice crystal.

Calculations were performed to determine the optimal truncation fraction $t$ which yields a crystal with the widest band gap. Initial calculations were performed at a sphere-density of 200 $(1/l_y)$. These initial calculations were performed for $t$ ranging from $t = 0$ to 0.225, in steps of 0.025. Figure 9 shows that a maximum width of the band gap of around $(\frac{\Delta\omega}{\omega_{\text{gap}}}) = 23\%$ is found around $t = 0.1$. More precise calculations, with a sphere-density of 400 $(1/l_y)$, were performed around this value of $t$. The chosen range was $t = 0.08$ to 0.12 in steps of 0.01. The results, also indicated in Figure 9, show that a maximum width of the band gap of 23.6% is found at a truncation fraction of $t = 0.1 \pm 0.01$. At this truncation fraction the particles are perfect truncated tetrahedra and the volume fraction of the assembled crystal is around 21%, see appendix C. The calculated maximum width of the band gap of 23.6% compares favorably to other types of self-assembled diamond-lattice photonic crystals that are described in literature. For instance, for hard-spheres in a diamond crystal lattice band gaps of around 7% are reported.[12]

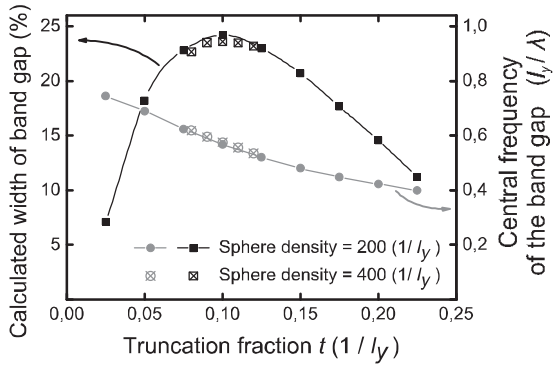

FIG. 9: Calculated band gap properties versus truncation fraction $t$. Initial calculations were performed with sphere-densities of 200 $(1/l_y)$ and more precise calculations with sphere-densities of 400 $(1/l_y)$. The black squares show how the calculated width of the band gap $(\frac{\Delta\omega}{\omega_{\text{gap}}})$ evolves with increasing truncation fraction $t$. A maximum width of the band gap of 23.6% is found at $t = 0.1 \pm 0.01$. The grey circles show the central frequency of the band gap $\omega_{\text{gap}}$ as a function of $t$. At the optimal position, $\omega_{\text{gap}} = 0.575\ (l_y/\lambda)$. With increasing $t$, the central frequencies are reduced.

Figure 9 also shows how the central frequency of the band gap $\omega_{\text{gap}}$ depends on the truncation fraction $t$. With increasing $t$, the central frequency is reduced. This is due to the fact that at larger truncation fractions $t$, the volume fraction of silicon is increased and therefore the average refractive index in the material is higher. From Bragg's law, adapted for photonic crystals, it can be deduced that this results in lower frequencies for the band gap.[32,35] At the optimal truncation fraction $t = 0.1$, the band gap has a calculated central frequency of 0.575 $(l_y/\lambda)$. Calculated for the target telecommunication wavelength $\lambda = 1550$ nm, this corresponds to a lattice spacing of $l_y = 891$ nm. The calculated band structure for an optimal crystal built from tetrahedra with $t = 0.1$ is shown in Figure 10.

When manufacturing the truncated tetrahedra, it is expected that the correct truncation fraction can be obtained with high accuracy. However, it must be realized that the achieved truncation fraction will never be $t = 0.1$ exactly. A small deviation from the perfect geometry is expected. For crystals built from these truncated tetrahedra, Figure 9 shows that even for large variations in the truncation fraction the calculated band gap that remains is very wide. In fact, a width of more than $(\frac{\Delta\omega}{\omega_{\mathrm{gap}}}) = 17.5\%$ remains for variations in the truncation fraction up to $\Delta t = 0.05$. It is expected that truncation of the tetrahedra can be achieved with a much higher precision than this $\Delta t$. Therefore we conclude that any uncertainties in the truncation step during the fabrication process have a negligible effect on the obtainable width of the band gap in the crystal structures. There will be a small change in the central frequency of the band gap, which needs to be taken into account.

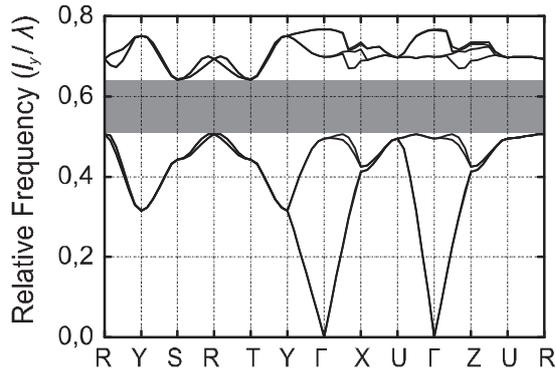

FIG. 10: *Mpb*-calculated bandstructure of a crystal with a diamond lattice built from perfect truncated tetrahedra ($t = 0.1$). The truncated tetrahedra have a dielectric constant of $\varepsilon_{\mathrm{Si}} = 12.25$. A wide photonic band gap that extends from $\omega = 0.507$ to $0.642$ ($l_y/\lambda$) is apparent (grey bar). This band gap has a central frequency of $\omega_{\mathrm{gap}} = 0.575$ ($l_y/\lambda$) and a width of $(\frac{\Delta\omega}{\omega_{\mathrm{gap}}}) = 23.6\%$.

A detailed description of the influence of other geometric imperfections on the width of the band gap of these assembled structures is beyond the scope of this paper. Nonetheless, a few remarks can be made. Existing literature shows for many types of diamond lattice photonic crystals that the band gap is very robust to withstand structural disorder. See for examples the results for inverse woodpile photonic crystals and for square spiral photonic crystals, described in references[30,31] and[36] respectively. This robustness appears to be more or less general for photonic crystals with a diamond-lattice.[10] An exception is encountered in literature: the band gap is very sensitive and closes rapidly when, due to disorder, loss of connectivity occurs in the dielectric material.[37] It is evident however, that loss of connectivity is not an issue for the structures described in this report. Firstly, it is not possible to assemble crystals without the truncated tetrahedral particles touching each other. Secondly, the large contact areas make sure that there is inter-particle contact even if slight misalignments occur. Thirdly, based on results described for rod-connected diamonds we can expect that even if particles are not connected to all four neighboring particles the band gap is not closed.[10] Thus, we expect a robustness to withstand disorder to be also present in the structures assembled from truncated tetrahedra that are described in this report. The calculations shown above for the truncation fraction are in agreement with this expectation. Nonetheless, the width of the band gap may be less if the periodicity of the crystal-lattice of the assembled structure is reduced due to a large polydispersity of the assembled particles. As an extreme example, it was shown that crystallization of hard spheres in an face centered cubic (fcc) crystal lattice no longer occurs when the polydispersity of the spheres exceeds a critical value of 12%.[38] In our case however, the polydispersity of the fabricated tetrahedral particles is expected to be very low due to the fact that they are micromachined on a single wafer under well-controlled conditions. Therefore, we do not expect a significant reduction in periodicity, and thus of the width of the band gap, due to polydispersity of the particles.

## V. CONCLUSIONS

In this report the optical properties of diamond lattice photonic crystals built from silicon tetrahedral and truncated tetrahedral particles were calculated. We have shown that crystal structures built from silicon tetrahedra are expected to display small stop gaps. By considering truncated tetrahedral particles as elements of diamond-lattice photonic crystals, the volume fraction is increased and photonic band gaps appear. It is shown that for a truncation fraction of 0.1 the volume fraction of the obtainable structures is increased from 4.17% to 21% and that the photonic band gap has a maximum width of the band gap. At this truncation fraction, the particles are equal to what in geometry is known as truncated tetrahedrons, *i.e.*, all edges of the particle have the same length. Furthermore, it was shown that the width of the band gap is insensitive to small deviations from the optimal truncation fraction that may occur during the fabrication of these particles.

A photonic crystal built from these elements has a width of the band gap of 23.6%, which is much wider compared to other types of self-assembled diamond-lattice photonic crystals that are described in literature. We envision three-dimensional self-assembly of these photonic crystals. The extra contact surfaces that arise because of truncation can be functionalized to assist the self-assembly process. The calculated widths of stop gaps and photonic band gaps can be used as reference values for optical reflectivity experiments to judge the quality of obtained crystal structures.

In conclusion, a novel class of diamond-lattice photonic band gap crystals can be obtained from silicon truncated tetrahedra. Since we aim to obtain these crystals through self-assembly, it is an exciting prospect to investigate the subsequent integration of these strongly photonic structures into existing silicon microfluidics and -electronics.

**Acknowledgements**

The authors acknowledge Erwin Berenschot and Johan Engelen for helpful discussions. Thijs Bolhuis and Henk van Wolferen are acknowledged for their assistance in setting up the desktop computer on which the *mpb*-calculations were performed.

**Appendix A**

The following space-filling procedure was used to obtain the unit-cells used in *mpb*:

- The coordinates of the vertices of a tetrahedron of type *I* and one of type *II* were determined in a suitable box.

- Using these coordinates, the planes that form the boundaries of the tetrahedra were defined.

- With a set resolution, ($x$,$y$,$z$)-coordinates of points inside and on the faces of the tetrahedra were determined and collected in an array for tetrahedron type *I* and an array for type *II*.

- These collected coordinates were set as the centers of the space-filling spheres. The radius of the spheres depends on the set resolution and was set at a value slightly larger than necessary to avoid interfacing issues (+5%).

- With lattice spacing $l_y$ set as 1, the size and axis of the tetrahedra were scaled to the correct dimensions in the unit-cells. Now the spheres are distributed in an orthorhombic lattice with a $(1/\sqrt{2})$ ratio along the $x$- and $z$-axis with respect to the $y$-axis.

- To obtain the required unit-cell, the four tetrahedra (2x*I*, 2x*II*) are placed at the correct positions in the unit-cell by $x$-,$y$-, and $z$-translations of the coordinates of the centers of the spheres.

- Truncated tetrahedra are obtained as follows: the size of the tetrahedra is increased whilst adding extra planes at the original vertices to define the truncated faces. Truncation is discussed in more detail in section IV B.

- With the arrays of coordinates and the radii of the spheres set, files were written with output suitable for the *mpb*-program and the ray-tracing program that was used to inspect resulting geometries.

**Appendix B**

The volume fraction, or filling fraction, of silicon $\phi_{Si}$ in a diamond-lattice crystal built from silicon tetrahedra is calculated as follows:

The volume of an untruncated tetrahedron $V_{ut}$ is given by

$$V_{ut} = \frac{l_{ut}^3}{6\sqrt{2}} \qquad (4)$$

here $l_{ut}$ is the length of the edges of the tetrahedra. In this case

$$l_{ut} = \frac{l_{xz}}{2} = \frac{l_y}{2\sqrt{2}}. \qquad (5)$$

The volume of the orthorhombic unit-cell $V_{uc}$ is equal to

$$V_{uc} = l_y\, l_x\, l_z = l_y\, l_{xz}^2 = \frac{l_y^3}{2}. \qquad (6)$$

With these two equations the volume fraction $\phi$ of tetrahedron material in the orthorhombic unit-cell can be determined. There are four tetrahedra in the unit-cell, therefore the volume fraction is equal to

$$\phi_{Si} = \frac{4\, V_{ut}}{V_{uc}} = \frac{2\sqrt{2}}{3}\left(\frac{l_{ut}}{l_y}\right)^3. \qquad (7)$$

Combining equations 5 and 7 yields

$$\phi_{Si} = \frac{1}{24} \approx 4.17\%. \qquad (8)$$

**Appendix C**

The volume fraction, or filling fraction, of silicon $\phi_{Si}$ in a diamond-lattice crystal built from perfect truncated silicon tetrahedra is calculated in this appendix. The following relations are used in this calculation:

The edges $l_{ut}$ of original untruncated tetrahedrons in the unit-cell have lengths equal to:
$$l_{ut} = \frac{l_y}{2\sqrt{2}}. \tag{9}$$
Furthermore, in case of a perfect truncated tetrahedron the truncation fraction $t$ is equal to
$$t = \frac{l_y}{10}. \tag{10}$$
The volume of a perfect truncated tetrahedron $V_{tt}$ is described by the following equation
$$V_{tt} = \frac{23\sqrt{2}}{12} l_{tt}^3, \tag{11}$$
with $l_{tt}$ the length of the edges of the truncated tetrahedra.

The relation between the edges $l_{tt}$, $l_{ut}$, and the truncation fraction $t$ is as follows:
$$l_{tt} = l_{ut} - \sqrt{2}t. \tag{12}$$
Therefore, for perfect truncated tetrahedra:
$$l_{tt} = \frac{3\sqrt{2}}{20} l_y \tag{13}$$
The volume of the orthorhombic unit-cell is equal to
$$V_{uc} = l_y\, l_x\, l_z = l_y\, l_{xz}^2 = \frac{l_y^3}{2}. \tag{14}$$

Given that there are four perfect truncated tetrahedra in the unit-cell, the above equations combine into an expression for the the volume fraction $\phi$ in the orthorhombic unit-cell:
$$\phi_{Si} = \frac{4\, V_{tt}}{V_{uc}} = \frac{19872}{96000} \approx 20.7\%. \tag{15}$$
Therefore the filling fraction of silicon in a diamond lattice crystal assembled from perfect truncated tetrahedra is equal to $\phi_{Si} = 20.7\%$.

---

* Electronic address: l.a.woldering@utwente.nl